\documentstyle[aps]{revtex}
\date{}             
\begin{document}
\title{The $\eta NN$ coupling constant }
\author{Shi-Lin Zhu\\
Department of Physics, University of Connecticut, U-3046\\ 
2152 Hillside Road, Storrs,  CT 06269-3046}
\maketitle
\begin{center}
\begin{minipage}{120mm}
\vskip 0.6in
\begin{center}{\bf Abstract}\end{center}
{\large
We derive the light cone QCD sum rule for the $\eta NN$ coupling constant
$g_{\eta NN} $. The contribution from the excited states and the continuum 
is subtracted cleanly through the double Borel transform with respect to 
the two external momenta, $p_1^2$, $p_2^2=(p-q)^2$. Our result is 
$\alpha_{\eta NN}=(0.3\pm 0.15)$, which favors small values used in 
literatures.

\vskip 0.5 true cm
PACS Indices: 14.40.Aq; 13.75.Gx; 13.75.Cs 
}
\end{minipage}
\end{center}

\large`
\section{Introduction}
\label{sec1}
Nowadays quantum chromodynamics (QCD) is widely believed to be the underlying
theory of the strong interaction. Yet the non-abelian nature of the gauge group 
makes analytical calculation extremely difficult in the low
energy sector. A typical example is the various coupling constants of 
meson nucleon interaction. These couplings are inputs for the one boson 
exchange potentials for the nuclear forces and the analysis of 
the important pseudoscalar and vector meson photo- and electro-production 
experiments currently underway in MAMI (Mainz) and Spring8 (JHF) etc.
For the pion nucleon sector there is enough precise data to extract these
couplings. Then they are used as inputs to make predictions and 
analyze other experimental data. In the kaon nucleon hyperon sector the 
situation is not so encouraging. But there is still some data available. 
The worst occurs in the $\eta NN$ and $\eta' NN$ sector, where knowledge 
of them is rather poor. In the present paper we shall focus on the 
calculation of $\eta NN$ coupling constant.

There were some theoretical papers on this issue. But the results from 
various approaches differed greatly. With $SU_f(3)$ symmetry it was 
found $\alpha_{\eta NN}={g^2_{\eta NN}\over 4\pi}=3.68$ 
from the analysis of the nucleon nucleon potential \cite{su3}.
Similar values was obtained in the non-relativistic model \cite{quark}.
From the analysis of forward nucleon nucleon scattering using the 
dispersion relation it was found that $\alpha_{\eta NN}< 1$, 
consistent to be zero \cite{grein}. 
In \cite{hat} the author was able to relate the proton matrix element
of flavor singlet current in the large $N_c$ limit to the pseudoscalar 
meson nucleon coupling constants, leading to $\alpha_{\eta NN} \sim 1.3$.
Typical values of $\alpha_{\eta NN}$ obtained in fits with one boson
exchange potentials range from 3 to 7 since the eta meson does not 
contribute significantly to the $NN$ phase shifts and nuclear binding 
at normal densities \cite{obep}. However this coupling is smaller than 1
and can be neglected in the full Bonn potential \cite{bonn}.
$\alpha_{\eta NN}$ extracted from the reaction $\pi^- p \to \eta n$
lies between 0.6--1.7 \cite{peng}. An interesting indirect constraint of $\alpha_{\eta NN}$ 
comes from the $\pi$-$\eta$ mixing amplitude generated by $\bar N N$ loops 
and neutron proton mass difference using hadronic models. In order to let 
this amplitude agree with results from chiral perturbation theory, 
$\alpha_{\eta NN}$ is required to be in the range $0.32$-$0.53$ \cite{chpt}. 
Eta meson photo-production did not fix $\alpha_{\eta NN}$ either. In \cite{prl}
$\alpha_{\eta NN}$ was suggested to around $1.0$ or $1.4$. Yet a recent
analysis of more precise eta meson photo-production experiments in Mainz suggested 
smaller value of $\alpha_{\eta NN}$ \cite{tiator}. 
In other words, the eta nucleon coupling constant is still very controversial. 
To derive it within an independent and reliable theoretical framework shall 
prove valuable. We shall use the now well developed light cone QCD sum rules (LCQSR) 
technique to calculate $\alpha_{\eta NN}$ in this work. Note our approach 
differs from all the above ones in that it starts microscopically from 
the QCD Lagrangian.

QCD sum rules (QSR) \cite{SVZ} are successful when applied to 
the low-lying hadron masses and couplings. In this approach 
the nonperturbative effects are introduced via various condensates in the vacuum. 
The light cone QCD sum rule differs from the conventional short-distance QSR in that 
it is based on the expansion over the twists of the operators. The main contribution
comes from the lowest twist operator. Matrix elements of nonlocal operators 
sandwiched between a hadronic state and the vacuum defines the hadron wave
functions. When the LCQSR is used to calculate the coupling constant, the 
double Borel transformation is always invoked so that the excited states and 
the continuum contribution can be subtracted quite cleanly. Moreover, the final 
sum rule depends only on the value of the hadron wave function at  
the middle point $u_0={1/2}$ for the diagonal case, which is much better 
known than the whole wave function \cite{bely95}. 
In the present case our sum rules involve with the eta wave function (EWF)
$\varphi_{\eta}(u_0 ={1\over 2})$ etc. These parameters 
are universal in all processes at a given scale.

We have used QCD sum rules to study the meson nucleon strong interactions. 
In \cite{pi1} the pion is treated as the external field to analyze the possible 
isospin symmetry violations of the pion nucleon coupling constant. Later 
the light cone QSR (LCQSR) was employed to extract the $\pi N N(1535)$ coupling constant,
which was found to be strongly suppressed \cite{pi2}. The same formalism was 
extended to the case of vector meson nucleon interaction \cite{vector}. 
The values of the vector and tensor coupling constants and their ratios of 
$\rho NN$ and $\omega NN$ interaction from the LCQSR agree well with the 
ones from the experimental data and the dispersion relation analysis. With 
the advent of the eta meson distribution amplitudes up to twist four \cite{ball}, we 
are now able to calculate the $\eta NN$ coupling constant with a theoretically 
well developed formalism. Although $\eta$ meson is a Goldstone boson, its mass
is not small in the real world and comparable with the typical hadronic scale 
due to the explicit breaking of $SU_f(3)$ flavor symmetry. 
We have included the eta mass correction in our calculation. 
Moreover the eta meson is an isoscalar, which leads to the big difference of 
the LCQSR for the $\eta NN$ coupling constant from that for the $\pi^0 NN$
coupling. We arrive at $\alpha_{\eta NN}={g^2_{\eta NN}\over 4\pi}=(0.3\pm 0.15)$.
The numerically small value is due to the cancellation between the leading term
and mass correction terms. This point can be seen clearly in later sections.

Our paper is organized as follows:
Section \ref{sec1} is an introduction.
We introduce the two point function for the $\eta NN$
vertex and saturate it with nucleon intermediate states in section \ref{sec2}. 
The definitions of the eta wave functions (EWF) are also presented.
Numerical analysis and a short summary is given in the last section.

\section{The LCQSR for the $\eta NN$ coupling}
\label{sec2}
We start with the two point function 
\begin{equation}
\Pi (p_1,p_2,q) = i\int d^4 x e^{ip x} 
\left \langle 0 \vert {\cal T}
\eta_p (x)  {\bar{\eta_p}} (0) \vert \eta (q) \right \rangle
\label{three-point}
\end{equation}
with $p_1 =p$, $p_2 = p-q$ and 
the Ioffe's nucleon interpolating field \cite{IOFFE}
\begin{equation}\label{cur1}
\eta_p (x) = \epsilon_{abc}  [  u^a (x) {\cal C} \gamma_\mu
u^b (x) ] \gamma_5 \gamma^\mu d^c (x) \; ,
\label{eq4}
\end{equation}
\begin{equation}
{\bar\eta}_p(y) = \epsilon_{abc}[{\bar u}^b(y) \gamma_\nu C 
         {\bar u}^{aT}(y) ] {\bar d}^c(y) \gamma^\nu \gamma^5\; ,
\label{eq5}
\end{equation}
where $a,b,c$ is the color indices
and ${\cal C} = i \gamma_2 \gamma_0$ is the charge conjugation matrix.
For the neutron interpolating field, $u \leftrightarrow d$. 

$\Pi (p_1,p_2,q)$ has the general form
\begin{equation}\label{eq3}
\Pi (p_1,p_2,q) =
 F (p_1 ^2 , p_2 ^2 ,q^2) {\hat q} \gamma_5 +
F_1 (p_1 ^2 , p_2 ^2 ,q^2) \gamma_5 + 
F_2 (p_1 ^2 , p_2 ^2 ,q^2) {\hat p} \gamma_5 +
F_3 (p_1 ^2 , p_2 ^2 ,q^2) \sigma_{\mu\nu}  \gamma_5 p^\mu q^\nu
\end{equation}

The sum rules derived from the chiral even tensor 
structure yield better results than those from the chiral even ones
in the QSR analysis of the nucleon mass \cite{IOFFE}. We shall focus on the 
tensor structure ${\hat q} \gamma_5$ and study the function $F(p_1^2, p_2^2, q^2)$ 
as in the QSR analysis of the pion nucleon coupling constant.

The eta nucleon coupling constant $g_{\eta NN}$ is defined by the
$\eta N$ interaction Lagrangian:
\begin{equation}
{\cal L}_{\eta NN} =  g_{\eta NN} {\bar N} i \gamma_5 \eta N 
.\; .
\label{eq5a}
\end{equation}

At the phenomenological level the eq.(\ref{three-point}) can be expressed as:
\begin{equation}\label{pole}
\Pi (p_1,p_2,q)  =   i \lambda_N^2 m_N g_{\eta NN} (q^2)
{  \gamma_5 {\hat q}
\over{ (p_1 ^2 - M_N ^2) (p_2 ^2 - M_N ^2) }} +\cdots
\label{phen1}
\end{equation}
where we include only the tensor structure $\gamma_5 {\hat q}$ only. 
The ellipse denotes the continuum and the single pole excited states to nucleon
transition contribution. $\lambda_N$ is the overlapping amplitude of 
the interpolating current $\eta_N (x)$ with the nucleon state
\begin{equation}
\left \langle 0 \vert \eta_N (0) \vert N (p) \right \rangle 
= \lambda_N  u_N (p)
\end{equation}

Neglecting the four particle component of the eta wave function, 	
the expression for $F(p^2_1,p^2_2,q^2)$ 
with the tensor structure at the quark level reads,
\begin{eqnarray}\label{quark}\nonumber
&&i\int d^4x e^{ipx} 
\langle 0| T\eta_p(x) {\bar \eta}_p(0) |\eta (q)\rangle =\\ \nonumber
&&2{\rm i}\int d^4x e^{ipx} \epsilon^{abc}\epsilon^{a^{\prime} b^{\prime} c^{\prime}} 
Tr \{ \gamma_\nu C {iS^T}_u^{b b^{\prime}}(x) C\gamma_\mu iS_u^{a a^{\prime}}(x)\}
\gamma_5 \gamma_\mu \langle 0| d^c (x) {\bar d}^{c^{\prime}}(0)|\eta  (q)\rangle
\gamma_\nu \gamma_5\\
&&+
4{\rm i}\int d^4x e^{ipx} \epsilon^{abc}\epsilon^{a^{\prime} b^{\prime} c^{\prime}} 
Tr \{ \gamma_\nu C {iS^T}_u^{b b^{\prime}}(x) C\gamma_\mu 
\langle 0| u^a (x) {\bar u}^{a^{\prime}}(0)|\eta  (q)\rangle\}
\gamma_5 \gamma_\mu iS_d^{c c^{\prime}}(x) \gamma_\nu \gamma_5
\end{eqnarray}
where $iS(x)$ is the full light quark propagator with both perturbative  
term and contribution from vacuum fields \cite{pi2}.

By the operator expansion on the light-cone
the matrix element of the nonlocal operators between the vacuum and 
eta state defines the two and three particle eta wave function. In order 
to simplify the notations we use ${\bar q \Gamma_\mu q}$ to denote 
$({\bar u \Gamma_\mu u}+ {\bar d \Gamma_\mu d} -2 {\bar s \Gamma_\mu s})/ \sqrt{6} $.
We also introduce $F_\eta ={f_\eta \over \sqrt{6}}$, where $f_\eta $ is defined as 
\begin{equation}
<0|\bar q (0)\gamma_\mu \gamma_5 q(0)| \eta (q) > =i f_\eta q_\mu \; .
\end{equation}

Up to twist four the Dirac components of this wave function can be 
written as \cite{ball}:

\begin{eqnarray}\label{phipi}\nonumber
&&<0| {\bar q} (0) \gamma_{\mu} \gamma_5 q(x) |\eta (q)>=i f_{\eta} q_{\mu} 
\int_0^1 du \; e^{-iuqx} [\varphi_{\eta}(u) 
+{1\over 16} m^2_\eta x^2 A(u)  ] \\ &&
+{i\over 2}f_\eta m_\eta^2 {q_\mu \over q x} 
\int_0^1 du \; e^{-iuqx}  B(u) + O(x^4) \; ,
\end{eqnarray}

\begin{equation}\label{phi_P}
<0| {\bar q} (0) i \gamma_5 q(x) |0>=
f_\eta \mu_\eta \int_0^1 du \; e^{-iuqx} \varphi_P (u)
\; ,
\end{equation}

\begin{equation}\label{phi_sigma}
<0| {\bar q} (0) \sigma_{\mu \nu} \gamma_5 q(x) |0>=
{i\over 6} f_\eta \mu_\eta (q_\mu x_\nu-q_\nu x_\mu) 
\int_0^1 du \; e^{-iuqx} \varphi_\sigma (u) \; ,
\end{equation}

\begin{eqnarray}
& &<0 | {\bar q} (0) \sigma_{\alpha \beta} \gamma_5 g_s 
G_{\mu \nu}(ux) q(x) |\eta (q)>=\nonumber \\ &&i f_\eta \mu_\eta \eta_3 
[(q_\mu q_\alpha g_{\nu \beta}-q_\nu q_\alpha g_{\mu \beta})
-(q_\mu q_\beta g_{\nu \alpha}-q_\nu q_\beta g_{\mu \alpha})]
\int {\cal D}\alpha_i \; 
\varphi_{3 \eta} (\alpha_i) e^{-iqx(\alpha_1+v \alpha_3)} \;\;\; ,
\label{p3pi} 
\end{eqnarray}

\begin{eqnarray}
& &<0| {\bar q} (0) \gamma_{\mu} \gamma_5 g_s 
G_{\alpha \beta}(vx) q(x) |\eta (q) >=
\nonumber \\
&&f_{\eta}m_\eta^2 \Big[ q_{\beta} \Big( g_{\alpha \mu}-{x_{\alpha}q_{\mu} \over q \cdot 
x} \Big) -q_{\alpha} \Big( g_{\beta \mu}-{x_{\beta}q_{\mu} \over q \cdot x} 
\Big) \Big] \int {\cal{D}} \alpha_i \varphi_{\bot}(\alpha_i) 
e^{-iqx(\alpha_1 +v \alpha_3)}\nonumber \\
&&+f_{\eta} m_\eta^2 {q_{\mu} \over q \cdot x } (q_{\alpha} x_{\beta}-q_{\beta} 
x_{\alpha}) \int {\cal{D}} \alpha_i \varphi_{\|} (\alpha_i) 
e^{-iqx(\alpha_1 +v \alpha_3)} \hskip 3 pt  \label{gi} 
\end{eqnarray}
\noindent and
\begin{eqnarray}
& &<0| {\bar q} (0) \gamma_{\mu}  g_s \tilde G_{\alpha \beta}(vx)q(x) |\eta (q)>=
\nonumber \\
&&-i f_{\eta} m_\eta^2 
\Big[ q_{\beta} \Big( g_{\alpha \mu}-{x_{\alpha}q_{\mu} \over q \cdot 
x} \Big) -q_{\alpha} \Big( g_{\beta \mu}-{x_{\beta}q_{\mu} \over q \cdot x} 
\Big) \Big] \int {\cal{D}} \alpha_i \tilde \varphi_{\bot}(\alpha_i) 
e^{-iqx(\alpha_1 +v \alpha_3)}\nonumber \\
&&-i f_{\eta} m_\eta^2  {q_{\mu} \over q \cdot x } (q_{\alpha} x_{\beta}-q_{\beta} 
x_{\alpha}) \int {\cal{D}} \alpha_i \tilde \varphi_{\|} (\alpha_i) 
e^{-iqx(\alpha_1 +v \alpha_3)} \hskip 3 pt . \label{git} 
\end{eqnarray}
\noindent 
The operator $\tilde G_{\alpha \beta}$  is the dual of $G_{\alpha \beta}$:
$\tilde G_{\alpha \beta}= {1\over 2} \epsilon_{\alpha \beta \delta \rho} 
G^{\delta \rho} $; ${\cal{D}} \alpha_i$ is defined as 
${\cal{D}} \alpha_i =d \alpha_1 
d \alpha_2 d \alpha_3 \delta(1-\alpha_1 -\alpha_2 
-\alpha_3)$. 
Due to the choice of the
gauge  $x^\mu A_\mu(x) =0$, the path-ordered gauge factor
$P \exp\big(i g_s \int_0^1 du x^\mu A_\mu(u x) \big)$ has been omitted.

The EWF $\varphi_{\eta}(u)$ is of twist two ,  
$\varphi_P(u)$, $\varphi_\sigma (u)$, and $\varphi_{3 \eta}$ are of twist three, 
while $A (u)$, part of $B(u)$ and all the EWFs appearing in 
eqs.(\ref{gi}), (\ref{git}) are of twist four.
The EWFs $\varphi (x_i,\mu)$ ($\mu$ is the renormalization point) 
describe the distribution in longitudinal momenta inside the eta meson, the 
parameters $x_i$ ($\sum_i x_i=1$) 
representing the fractions of the longitudinal momentum carried 
by the quark, the antiquark and gluon.

The normalization and definitions of the various constants can be found in 
\cite{ball}. Some of them are 
$\int_0^1 du \; \varphi_\eta(u)=\int_0^1 du \; \varphi_\sigma(u)=1$,
$\int {\cal D} \alpha_i \varphi_\bot(\alpha_i)=
\int {\cal D} \alpha_i \varphi_{\|}(\alpha_i)=0$ etc.

Since the steps to derive LCQSRs are very similar to those in \cite{pi2,vector}, 
we present final sum rule directly. Interested readers may consult the above 
papers for details.

\begin{eqnarray}\label{quark0}\nonumber
m_N \lambda^2_N  g_{\eta NN}e^{-{ M_N^2\over M^2} } = &\\ \nonumber
-e^{-{u_0(1-u_0)m^2_\eta \over M^2}} \{
-{F^u_\eta\over 2\pi^2} \varphi_\eta (u_0) M^6 f_2 ({s_0\over M^2})
+{m_\eta^2\over 4\pi^2} 2u_0 [F^u_\eta A(u_0) +F^d_\eta \phi_B (u_0)]
M^4 f_1 ({s_0\over M^2})  
&\\ \nonumber
-{F^u_\eta\over 9\pi^2} a\mu_\eta [\varphi_\sigma (u_0) 
+{u_0\over 2}\varphi_\sigma^\prime (u_0)] M^2 f_0 ({s_0\over M^2}) 
&\\ \nonumber
+{1\over 12\pi^2} F^u_\eta \mu_\eta \eta_3 a m_\eta^2 I_1 [\varphi_{3\eta}]
-{1\over 6\pi^2} F^u_\eta \mu_\eta \eta_3 a  I_2 [\varphi_{3\eta}]
M^2 f_0 ({s_0\over M^2}) 
&\\ \nonumber
+{1\over 2\pi^2} m_\eta^2 (F^u_\eta +F^d_\eta) 
\{ {1\over 4} I_2 [\varphi_{\|}] -{1\over 4} I_2 [\varphi_{\bot}] 
+I_1 [\varphi_{\|}] - I_4 [{\tilde \varphi}_{\|}]
-{1\over 4} I_7 [{\tilde \varphi}_{\bot}] \}
M^4 f_1 ({s_0\over M^2})
&\\
+{1\over 2\pi^2} m_\eta^4 (F^u_\eta +F^d_\eta) 
\{ - I_3 [\varphi_{\|}] - I_3 [\varphi_{\bot}] 
- I_5 [{\tilde \varphi}_{\|}] - I_6 [{\tilde \varphi}_{\|}]
+ I_5 [{\tilde \varphi}_{\bot}]+ I_6 [{\tilde \varphi}_{\bot}] \}
M^2 f_0 ({s_0\over M^2})
\}&\; ,
\end{eqnarray}
where
$f_n(x)=1-e^{-x}\sum\limits_{k=0}^{n}{x^k\over k!}$ is the factor used 
to subtract the continuum, $s_0$ is the continuum threshold.
$u_0={M^2_1 \over M^2_1 + M^2_2}$, 
$M^2\equiv {M^2_1M^2_2\over M^2_1+M^2_2}$,  
$M^2_1$, $M^2_2$ are the Borel parameters, 
and $\varphi_\sigma^\prime (u_0)  ={d\varphi_\sigma (u)\over du}|_{u=u_0}$.
In order to make comparison with the sum rule for $\pi^0 NN$ 
coupling constant $g_{\pi NN}$, we have labeled the eta meson decay constant
$F_\eta$ with the flavor index. 

The functions $I_i [\varphi_{3\eta}]$ etc are defined as:
\begin{equation}
\phi_B (u_0) =-\int_0^{u_0} du B(u)
\; ,
\end{equation}

\begin{equation}
I_1 [F] =2u_0 \int_0^{u_0} d\alpha_1 \int_0^{1-u_0} d\alpha_2
{F(\alpha_1, \alpha_2, 1-\alpha_1-\alpha_2) \over (1-\alpha_1-\alpha_2)^2}
(1-2u_0+\alpha_1-\alpha_2) 
\; ,
\end{equation}

\begin{eqnarray}\nonumber
&&I_2 [F] =\int_0^{u_0} d\alpha_1 
{F(\alpha_1, 1-u_0, u_0-\alpha_1) \over u_0-\alpha_1}+
\int_0^{1-u_0} d\alpha_2 
{F(u_0,\alpha_2, 1-u_0-\alpha_2) \over 1-u_0-\alpha_2}\\
&&-2\int_0^{u_0} d\alpha_1 \int_0^{1-u_0} d\alpha_2
{F(\alpha_1, \alpha_2, 1-\alpha_1-\alpha_2) \over (1-\alpha_1-\alpha_2)^2}
\; ,
\end{eqnarray}

\begin{equation}
I_3 [F] =2u_0 \int_0^{u_0} d\alpha_1 \int_0^{1-u_0} d\alpha_2
F(\alpha_1, \alpha_2, 1-\alpha_1-\alpha_2) 
{(u_0 -\alpha_1) (1-u_0-\alpha_2)\over (1-\alpha_1-\alpha_2)^2}
\; ,
\end{equation}

\begin{equation}
I_4 [F] =2u_0 \int_0^{u_0} d\alpha_1 \int_0^{1-u_0} d\alpha_2
{F(\alpha_1, \alpha_2, 1-\alpha_1-\alpha_2) \over 1-\alpha_1-\alpha_2}
\; ,
\end{equation}

\begin{equation}
I_5 [F] =2u_0 \int_0^{u_0} d\alpha_1 \int_0^{1-u_0} d\alpha_2
F(\alpha_1, \alpha_2, 1-\alpha_1-\alpha_2) 
{u_0 -\alpha_1\over 1-\alpha_1-\alpha_2}
\; ,
\end{equation}

\begin{equation}
I_6 [F] =2u_0 \int_0^{u_0} d\alpha_1 \int_0^{u_0-\alpha_1} d\alpha_3
F(\alpha_1, 1-\alpha_1-\alpha_3, \alpha_3) 
\; ,
\end{equation}

\begin{equation}
I_7 [F] =2u_0 \{ \int_0^{u_0} d\alpha_1 
{F(\alpha_1, 1-u_0, u_0-\alpha_1) \over u_0-\alpha_1}-
\int_0^{1-u_0} d\alpha_2 
{F(u_0, \alpha_2, 1-u_0-\alpha_2) \over 1-u_0-\alpha_2} \}
\; ,
\end{equation}
where $F=\varphi_{3\eta}, \varphi_{\|}, \varphi_{\bot}, {\tilde \varphi}_{\|},
{\tilde \varphi}_{\bot}$.

\section{Discussion}
\label{sec6}

Since eta meson is an isoscalar, we have 
$F^u_\eta=F^d_\eta=F_\eta ={f_\eta\over \sqrt{6}}$. Replacing the $\eta$ index 
with $\pi$ and $F_\eta$ by $f_\pi$ in (\ref{quark0}), we recover the sum rule for $g_{\pi NN}$
\cite{pi2}. Note $f^u_\pi=-f^d_\pi =f_\pi$. In other words, the twist four terms
involved with three particle pion wave functions vanish due to isospin symmetry.
The first term in (\ref{quark0}) is the leading twist two term. 
The third term is of twist three and of the same sign 
as the leading term. The second term comes from two particle EWF and the remaining 
terms all come from three particle EWFs. Although they are of twist four except the 
fourth term, 
their contribution is greatly enhanced by the factor $m_\eta^2$ in contrast with 
$m_\pi^2$ in the $\pi NN$ coupling case. Moreover they are of the 
opposite sign as the leading twist two 
and three terms, which leads to strong cancellation.
In other words, large mass and isoscalar structure of eta meson 
causes $g_{\eta NN}$ to be much smaller than $g_{\pi NN}$. 

The sum rule (\ref{quark0}) is symmetric and diagonal, which requires 
the Borel parameters $M_1^2=M_2^2$, i.e, $u_0 ={1\over 2}$.
The working interval for analyzing the QCD sum rule (\ref{quark0}) 
is $0.9\mbox{GeV}^2 \leq M_B^2\leq 1.8\mbox{GeV}^2$, a standard choice 
for analyzing the various QCD sum rules associated with the nucleon. 
In order to diminish the uncertainty due to $\lambda_N$, we shall 
divide (\ref{quark0}) by the Ioffe's mass sum rule for the nucleon:
\begin{equation}\label{mass}
32\eta^4 \lambda_N^2 e^{-{ M_N^2\over M^2} }
=M^6 f_2 ({s_0\over M^2})+{b\over 4}M^2 f_0 ({s_0\over M^2})
+{4\over 3}a^2 -{a^2m_0^2\over 3M^2} \; .
\end{equation}

The various parameters which we adopt are 
$f_\eta =(0.133\pm 0.01)$ GeV \cite{eta},
$\eta_3=0.013$, 
$a=-4\pi^2 <0|\bar q q|0>=0.67\, \mbox{GeV}^3$, 
$\mu_\eta =2.13 $GeV \cite{ball} at the scale $\mu =1$GeV, 
$s_0=2.25$GeV$^2$, $m_N=0.938$GeV, 
$\lambda_N =0.026$GeV$^3$ \cite{IOFFE}.  

At $u_0={1\over2}$ the values of various eta meson wave functions are: 
$\varphi_\eta(u_0)=1.05$, 
$A(u_0)=4.14$, $\phi_B (u_0)=0$,
$\varphi_\sigma(u_0)=1.44$,
$\varphi^\prime_\sigma(u_0)=0$,  
$I_1 [\varphi_{3\eta}] =0$,
$I_1 [\varphi_{\|}] =0.026$,
$I_2 [\varphi_{3\eta}] =-0.9375$,
$I_2 [\varphi_{\|}] =0$,
$I_2 [\varphi_{\bot}] =0$,
$I_3 [\varphi_{\|}] =0$,
$I_3 [\varphi_{\bot}] =0$,
$I_4 [{\tilde \varphi}_{\|}]=-0.313$,
$I_5 [{\tilde \varphi}_{\|}]=-0.032$,
$I_5 [{\tilde \varphi}_{\bot}]=0.0396$,
$I_6 [{\tilde \varphi}_{\|}]=-0.052$,
$I_6 [{\tilde \varphi}_{\bot}]=0.044$,
$I_7 [{\tilde \varphi}_{\bot}]=0$
at $u_0 ={1\over 2}$ and $\mu =1$GeV.

The dependence on the Borel parameter $M^2$ of $g_{\eta NN}$ 
are shown in FIG 1 with $s_0=2.35, 2.25, 2.15$ GeV$^2$.
The final sum rule is stable in the working region of the Borel parameter $M^2$.
We obtain:
\begin{equation}\label{num}
g_{\eta NN}=(1.7\pm 0.3)\; .
\end{equation}

In the above numerical analysis we have used relatively large quark condensate
value $<\bar q q>=-(240\pm 10)^3 $MeV$^3$, which corresponds to $a=0.67$GeV$^3$.
In the literatures another value $<\bar q q>=-(225\pm 10)^3 $MeV$^3$
and $a=0.55$GeV$^3$ is also used. Since we are not able to know very precisely
the quark condensate value, we also present the variation of $g_{\eta NN}$ 
with $M^2, s_0$ with $a=0.55$GeV$^3$ in FIG 2. In this case we have,
\begin{equation}\label{num1}
g_{\eta NN}=(2.1\pm 0.3)\; .
\end{equation}

We have included the uncertainty due to the variation of the continuum 
threshold and the Borel parameter $M^2$ in (\ref{num}) and (\ref{num1}). 
In other words, only the errors arising from numerical analysis of 
the sum rule (\ref{quark0}) are considered. Other sources of 
uncertainty include: (1) the truncation of OPE on the light cone at 
the twist four operators. For example the four particle 
component of EWF is discarded explicitly; (2) the EWFs are estimated 
with QCD sum rule, which also induces some errors;
(3) the continuum model used in the subtraction of contribution from the 
higher resonances and continuum spectrum; (4) errors in $f_\eta$ etc. 

With all these uncertainties we arrive at 
\begin{equation}\label{final2}
\alpha_{\eta N N}=(0.3 \pm 0.15) \; .
\end{equation}

For the $\eta, \eta'$ sector instanton effects might be important.
It's well known that a large part of $\eta'$ mass comes from the $U_A(1)$ anomaly. 
Through $\eta -\eta'$ mixing instantons also affect eta meson mass and
decay constant $f_\eta$. Fortunately we know from phenomenological analysis 
that the $\eta -\eta'$ mixing angle is about $-20$ degrees \cite{eta}. 
So for eta meson such effects may be not so large as in the $\eta'$ channel.
Direct instantons favor strongly the scalar and pseudoscalar channel and 
might affect the mass sum rules for the mesons in these channels.
In our QCD sum rule analysis of eta NN coupling constant we have 
chosen the tensor structure ${\hat q} \gamma_5$. 
Moreover we have used the experimental values for $m_\eta, f_\eta$ as inputs
instead of invoking the eta meson mass sum rules to extract them. 
Hence the possible correction from instantons is expected to be 
relatively small.  

In short summary we have calculated the eta nucleon coupling constant 
with the light cone QCD sum rules. 
The continuum and the excited states contribution is subtracted rather 
cleanly through the double Borel transformation. Our approach differs from 
all the available methods in the extraction of $g_{\eta NN}$ and starts from 
the quark gluon level. So it is independent and more reliable to some extent.
Our result of $\alpha_{\eta N N}$ favors the small value. Except the 
nonrelativistic quark model and fits with one boson exchange potentials, 
other approaches tend to yield small values for $\alpha_{\eta N N}$.
However in such potentials the eta meson was treated as some effective 
degree of freedom to model other multi-meson correlations. Hence the eta meson
in these potentials can not be related to the real 
eta meson seen in the photo- or electro-production experiments in a simple way.
In other words, the $\alpha_{\eta N N}$ in these potentials may be not the same 
quantity as the coupling we have calculated. We hope our extraction of 
$\alpha_{\eta N N}$ can be used to analyze future eta meson photo- and 
electro-production experiments.


\newpage
{\bf Figure Captions}
\vspace{2ex}
\begin{center}
\begin{minipage}{130mm}
{\sf FIG 1.} \small{The sum rule for $g_{\eta NN}$ as a function of 
the Borel parameter $M^2$ with $a=0.67$GeV$^3$ and the continuum threshold 
$s_0 =2.35, 2.25, 2.15$GeV$^2$. 
}
\end{minipage}
\end{center}
\begin{center}
\begin{minipage}{130mm}
{\sf FIG 2.} \small{The same notations as in FIG 1 except $a=0.55$GeV$^3$.
}
\end{minipage}
\end{center}

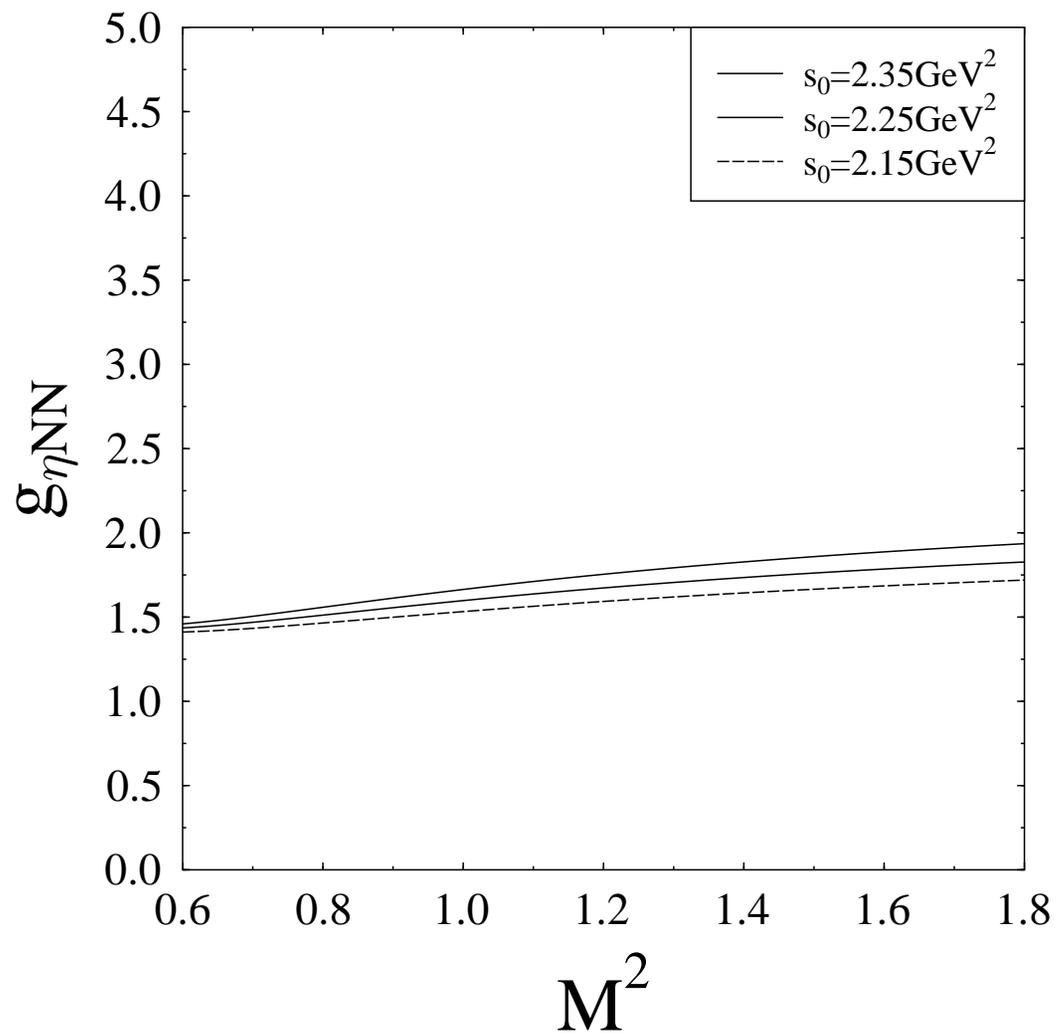
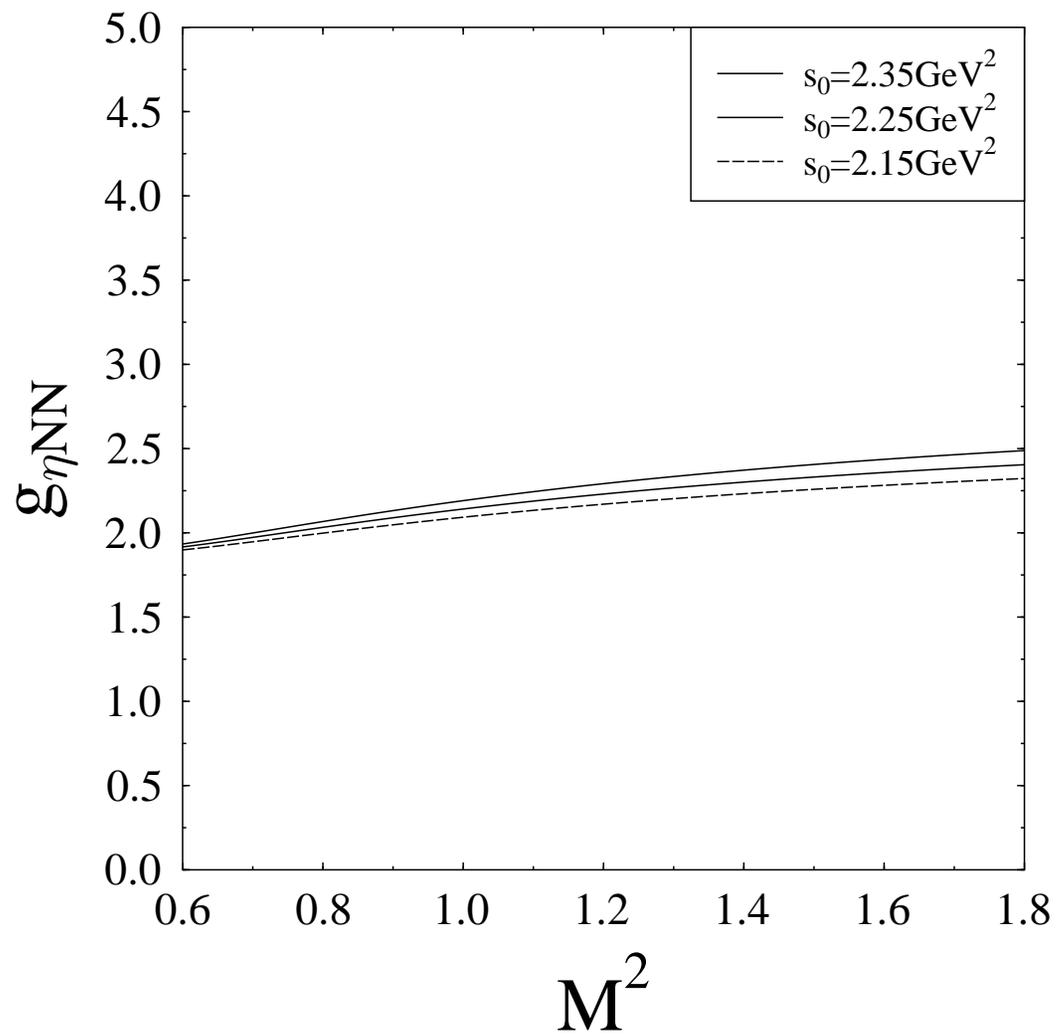
\end{document}